\documentclass{article}
\usepackage[utf8]{inputenc}
\usepackage{geometry}
 \usepackage{graphicx}
 \usepackage{titling}
 \usepackage[utf8]{inputenc}
 \usepackage{graphicx}
 \usepackage{listings}
 \usepackage{xcolor}
 \usepackage{multirow}
 \usepackage{hyperref}
 \usepackage{longtable}
 \usepackage{natbib}
 \usepackage{caption}
 \newsavebox{\mybigtable}
\title{Intelligence Preschool Education System based on Multimodal Interaction Systems and AI
}
\author{Long Xu}
\date{July 2024}
 
\usepackage{fancyhdr}
\fancypagestyle{plain}{
    \fancyhf{} 
    \fancyfoot[L]{\thedate}
    \fancyhead[L]{Description of Assignment}
    \fancyhead[R]{\theauthor}
}
\makeatletter
\def\@maketitle{%
  \newpage
  \null
  \vskip 1em%
  \begin{center}%
  \let \footnote \thanks
    {\LARGE \@title \par}%
    \vskip 1em%
  \end{center}%
  \par
  \vskip 1em}
\makeatother

\usepackage{lipsum}  
\usepackage{cmbright}
\usepackage{tocloft}
\begin{document}
\maketitle
\noindent\begin{tabular}{@{}ll}
    Author & \theauthor\\
\end{tabular}

\newpage  

\section{Introduction}
Rapid progress in AI technologies has generated considerable interest in their potential to address challenges in every field and education is no exception. Improving learning outcomes and providing relevant education to all have been dominant themes universally, both in the developed and developing world. And they have taken on greater significance in the current era of technology driven personalization.

\subsection{Research Motivation}
Although the Chinese government has invested heavily in education and has made great progress in all aspects, it still faces huge challenges in the field of preschool education.

Early childhood education is of great significance in fostering children's holistic development, forming habits, enhancing cognition, and reducing developmental gaps, etc.
However, the current early childhood education model has the following two shortcomings.
\begin{itemize}
    \item The quality of education varies widely across different regions, leading to educational inequity.
    \item The educational process is difficult to digitize, and the outcomes tend to rely more on subjective perception. This makes it challenging to promote effective educational methods and prevents the improvement of inadequate ones.
\end{itemize}

Inspired by the development of artificial intelligence and the internet, we propose a digital early childhood education system based on these technologies to address the first issue. For the second issue, we will focus on researching the key factors that influence the educational process (such as habit formation) and design digital solutions for these key elements. By creating a "perception-analysis-feedback" loop, we aim to achieve the digitization of early childhood education.

The purpose of this system is to provide more children with access to high-quality early childhood education opportunities and reduce the existing inequities. Simultaneously, we will leverage artificial intelligence and digital technologies to continuously improve the quality of education.

\subsection{What We Will Do}
In the following content, we will first analyze the application of AI in education, and further analyze the problems that exist in its practical application. Finally, we will propose our own solutions.

In addressing educational inequity, we focus on using internet technology and artificial intelligence to ensure that everyone has the same opportunity to receive high-quality education. In terms of digitizing the educational process, we will first define the key factors required for digital education and conduct research based on these factors. Throughout the process, multimodal human-computer interaction technologies, including \textbf{affective computing}, will be continuously applied to capture and influence the key elements of the digital education process.

We believe AI technology has the potential to help the world address 
these chronic challenges in education relating to equity, learning outcomes, and real world relevance. 

\newpage
\section{Children's Preschool Education and the Challenges It Faces}
\subsection{AI in Education}
\cite{baker2019educ} categorize AI in Education (AIED) applications into 3 categories: (i) learner-oriented AIED; (ii) instructor-oriented AIED; and (iii) institutional system-oriented AIED. However often applications that focus on learning cut cross several stakeholders especially in the current context and impact a common objective - learning. They are likely to reflect a collaboration among teachers, learners, parents, tutors and peers. We prefer the following modified categorization:
\begin{itemize}
    \item[(i)] learning-oriented, including learners, teachers, educational institutions, tutors, peers, parents and para educators;
    \item[(ii)] institutional operations oriented financial aid, retention, learner experience;
    \item[(iii)] policy oriented including governmental interventions, subsidies, competitive assessments like MCAS in the U.S.
\end{itemize}

\cite{guan2020artificial, SRINIVASAN2022100062} summarize the various definitions of AIED in prior research:
\begin{table}[!h]
\centering
\label{tab:current_situation}
\scalebox{0.83}{
\begin{tabular}{cl}
\hline
Authors & \multicolumn{1}{c}{Definition} \\ \hline
\cite{hwang2003conceptual} & \begin{tabular}[c]{@{}l@{}}Intelligent tutoring system that helps to organize system knowledge and operational\\ information to enhance operator performance and automatically determining exercise \\ progression and remediation during a training session according to past student performance.\end{tabular} \\ \hline
\cite{johnson2009intelligent} & \begin{tabular}[c]{@{}l@{}}Artificially intelligent tutors that construct responses in real-time using its own \\ ability to understand the problem and assess student analyses.\end{tabular} \\ \hline
\cite{popenici2017exploring} & \begin{tabular}[c]{@{}l@{}}Computing systems that are able to engage in human-like processes such as learning, \\ adapting, synthesizing, self-correction and use of data for complex processing tasks.\end{tabular} \\ \hline
\cite{chatterjee2020adoption} & \begin{tabular}[c]{@{}l@{}}Computing systems capable of engaging in human-like processes such as adapting, learning, \\ synthesizing, correcting and using of various data required for processing complex tasks.\end{tabular} \\ \hline
\end{tabular}
}
\end{table}

The first two definitions of AIED above are focused on intelligent tutoring and the last two are slightly broader but still focused on the learner.

\cite{srinivasan2022ai} mentioned, based on an analysis of over 400 research articles published between 2000 and 2019 on the application of AI for teaching and learning, they find shifts in research emphasis between the first and second decade. During the period 2000–2009, the emphasis was on learner-oriented approaches trending away from instructor-oriented approaches. In the second decade 2010–2019, they found that there was \textbf{a noticeable shift towards modeling learning outcomes/analytics and student profiling}. Their findings reflect the growing interest in explicit measurement of learning outcomes, and predictive AI models including the use of deep statistical learning.

Based on a review of 220 research articles on personalized language learning, \cite{chen2021twenty} find that (1) multimodal videos promote effective personalized feedback; (2) personalized context-aware ubiquitous language learning enables active interaction with the real world by applying authentic and social knowledge to their surroundings; (3) mobile chatbots with Automated Speech Recognition (ASR) provide human-like interactive learning experiences to practice speaking and pronunciation; (4) collaborative game-based learning with customized gameplay path and interaction and communication features supports social interactivity and learning of various language skills; (5) AI promotes effective outcome prediction and instruction adaptation for individuals based on massive learner data; (6) Learning Analytics (LA) dashboards facilitate personalized recommendations through learning data visualization; and (7) data driven learning allows personalized and immediate feedback in real-time practice. Generally, these findings validate that AI technology can be beneficial in personalized language learning.

The above reviews provide valuable summaries of the progress AIED has made, especially in personalized and adaptive learning. Broadly the findings support the hypothesis that personalization makes a difference when learner characteristics and differences are carefully considered. The trend towards more explicit measurement of learning outcomes is healthy and much needed. In incorporating learning styles, extant research with the exception of \cite{srinivasan2021improving} does not consider the beneficial impact of engaging multiple sensory channels. The extant research also does not explicitly propose a generalized framework for integrating findings from learning science. Granular personalization governed by learning styles including the influence of different sensory modes is integral to a more holistic AI-led learning framework.

\subsection{Current Limitations}
Analyzing the work of AI in education, we found that an effective AI learning system in early childhood education typically possesses the 7 characteristics mentioned by \cite{chen2021twenty}. However, most existing work tends to focus on only one aspect of these characteristics, and there is a lack of a medium to integrate these 7 features into a cohesive educational system.

This medium must have characteristics that allow it to be generated or processed by each link and be relevant to the educational process environment. Most importantly, this medium must have a certain degree of manipulability, so that by manipulating it, the educational process can be influenced.

Through investigation and analysis, we found that research on the mechanisms of habit formation (\cite{lewis2021shaping, lewis2021habit}) may reveal an excellent medium suitable for AI in education, which is the factors influencing habit formation. These factors not only influence habit formation but are essentially key elements affecting learning behavior. Moreover, these factors can be quantified, observed, and altered, making them perfectly applicable to research on AI in preschool education.

I believe the key here is to understand how to extract the critical factors that might influence or drive educational behavior from the three interactive scenarios of "child-teacher", "child-environment", and "teacher-environment".
By analyzing and optimizing these critical factors, we can enhance the success rate of educational behavior.

\newpage
\section{Hardware and software system}

Human-computer interaction is a highly effective method for extracting information and affective computing.
In our interactive education system, the hardware system is primarily used for students or teachers to interact with the backend algorithms. The software system mainly processes these interaction data, while also providing feedback to the teaching process by adjusting the parameters of the hardware system.

\subsection{Hardware Interaction System }
\begin{figure}[!h]
    \centering
    \includegraphics[width=1\linewidth]{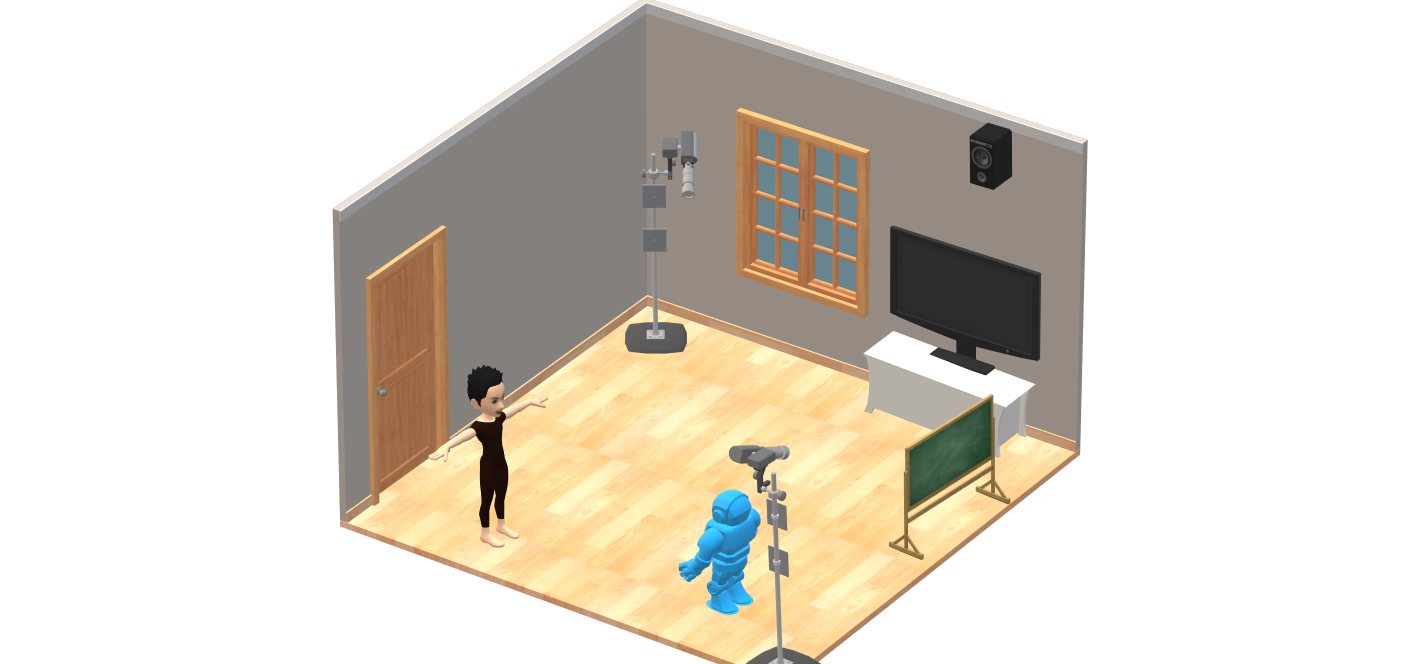}
    \caption{The hardware framework of our multimodal interaction system (for demonstration purposes only, see Table \ref{tab:hardware_list} for the quantity and specifications of the equipment)}
    \label{fig:hardware_env}
\end{figure}

To efficiently extract key factors in human-computer interaction, we first designed a hardware system for use in school or home settings, as shown in Figure \ref{fig:hardware_env}.

In our initial concept for the hardware system, it includes audio receivers, speakers, monitor cameras, computer, tablets, and laser projector, etc.

These devices can be used to capture interactions between children and teachers and perform related affective computing and educational behavior analysis. They can also directly interact with teachers or children, conduct data analysis, and provide status feedback.

In the initial phase of our research, we chose to use the free facilities provided by AIRS to set up the hardware system. Subsequent research will be based on this hardware. These free facilities can significantly reduce the cost of the hardware system and improve the efficiency of system setup.

\subsection{Software Interaction System }
Based on the aforementioned hardware system setup, we will also develop the corresponding software system.

In our software system, artificial intelligence algorithms play an important role.
Our software system can provide rich multimodal interaction capabilities. With the support of AI algorithms, our software system can perform real-time analysis and collection of teaching data, as well as provide guidance and interventions in the educational process.

\begin{figure}[!h]
    \centering
    \includegraphics[width=0.8\linewidth]{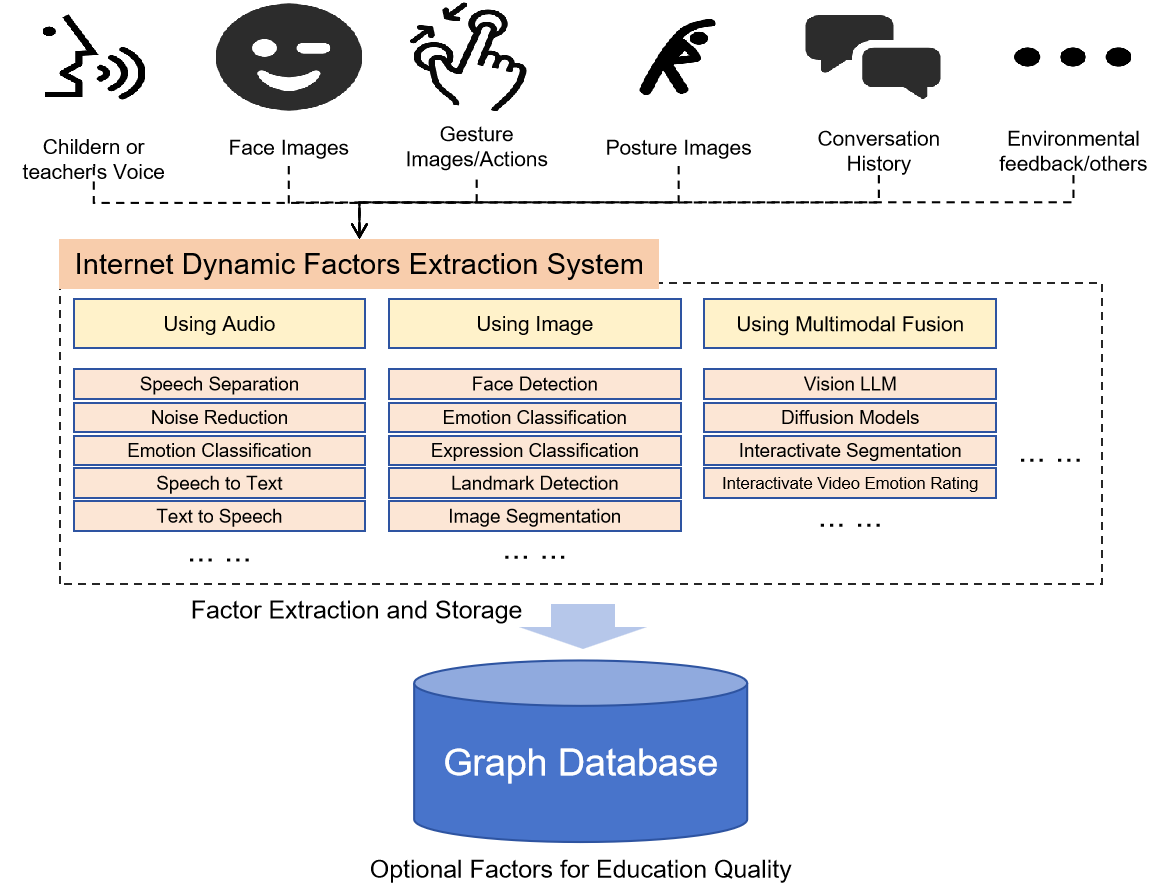}
    \caption{Diagram of the proposed multimodal human-computer interaction software system framework}
    \label{fig:overall_software_framework}
\end{figure}

The overall software framework are illustrated in Figure \ref{fig:overall_software_framework}. 
In our software system, in addition to developing some AI algorithms ourselves, we have also procured some mature technologies provided by third parties. For specific details, please refer to Table \ref{tab:softwarecost}.

This software system models the digital education process around multimodal input through methods such as affective computing, computer vision, machine learning, and data analysis.

Additionally, we use large model technology to provide realistic dialogue interactions, enhancing the teaching process's fun by offering children anthropomorphic interaction methods, thereby improving the quality of interaction and data.

This system is designed around the extraction, analysis, and feedback of factors during the interaction process, forming a complete digital education system. 

\newpage
\section{Extracting Factors Related to Education in Human-Computer Interaction (HCI)}

Before analyzing the educational process, we need to identify the key factors that influence it, including children's emotions, knowledge retention, teachers' emotions, teaching methods, and environmental temperature, among others.

To achieve this goal, we first need to design a user-friendly multimodal interaction system. The design principles of these systems should ensure that they are appealing to children, teachers, and other participants.
\cite{van2012further} and \cite{mclay2017acquisition} pointed that children with autism spectrum disorders (ASD) show a preference for one of the three communication modes: \textbf{manual signing}, \textbf{picture exchange}, and \textbf{speech-generating devices}. 
In addition, \cite{anderson1994interactive} pointed that successful communication involves active involvement of both participants, asking and answering questions, volunteering information, and responding sensitively to contributions from their partners.

Based on these research findings, we discovered that an interactive system appealing to children must first exhibit behavior patterns similar to those of humans. 
This includes voice conversations, asking questions, answering questions, providing additional information, and responding to children's emotional changes. 

Based on our designed multimodal interaction system of hardware and software, this section introduces several possible methods for interactive factor extraction.

\subsection{Extraction of Key Factors Based on Affective Computing}
Emotion is generally considered an important factor influencing education, and this factor can be conveniently implemented using facial recognition.

The research content of affective computing primarily involves the following five aspects:
\begin{itemize}
    \item Fundamental Theory
    \item Emotion Signal Collection
    \item Multimodal Integration
    \item Emotion Generation and Expression
    \item Emotion Analysis
\end{itemize}

In fundamental theoretical research, emotion models are divided into two types: discrete emotion models and dimensional emotion models. In our research plan, we use the dimensional emotion model for emotion analysis because the data in this modality is easier to quantify and can be more readily processed by AI algorithms.

In the current field of affective computing, data from sensors such as cameras and microphones are primarily used. Additionally, \textbf{physiological data}, which straightforwardly and objectively reflect an individual's state, have garnered increasing attention from researchers in recent years. Currently, commonly used physiological data include \textbf{electroencephalogram} (EEG), \textbf{electrodermal activity} (EDA), \textbf{respiration}, \textbf{skin temperature}, \textbf{electrocardiogram} (ECG), \textbf{electromyogram} (EMG), \textbf{blood volume pulse} (BVP), and \textbf{electrooculogram} (EOG). However, due to the impact of cost and the manner of wearing the sensors, the amount of such data is relatively limited in current research.

\newpage
\section{Identifying Important Factors in Education Process}
\label{chapter5}
The habit formation is somewhat related to the educational process. For example, both have a goal, and both are influenced by certain key factors. Initially inspired, we found that we can use algorithms that identify important factors relevant to the field of habit formation to pinpoint factors that significantly impact education.

\newpage
\section{Regulating Children's Behavior Using Important Factors}
\label{chapter6}
To regulate children's behavior using important factors, we first need to understand the basic principles of behavior control (\cite{frings2020binding}). Traditional theories suggest that behavior is controlled by two distinct systems: \textbf{the goal-directed system} and \textbf{the habit system}.

The goal-directed system refers to the system that makes decisions and takes actions based on the consequences of behavior. This system requires individuals to have a clear understanding of the relationship between behavior and its outcomes to make appropriate choices. Goal-directed behavior primarily involves the brain's prefrontal cortex and dorsomedial striatum (DMS). These areas are responsible for integrating information, planning, and decision-making. Therefore, the goal-directed system is more suited to using reinforcement learning. One classic experiment is the reward devaluation experiment, where animals reduce behaviors related to rewards that are no longer attractive (for example, when food is poisoned).

\bibliographystyle{plainnat}
\bibliography{ref}

\begin{thebibliography}{16}
\providecommand{\natexlab}[1]{#1}
\providecommand{\url}[1]{\texttt{#1}}
\expandafter\ifx\csname urlstyle\endcsname\relax
  \providecommand{\doi}[1]{doi: #1}\else
  \providecommand{\doi}{doi: \begingroup \urlstyle{rm}\Url}\fi

\bibitem[Anderson et~al.(1994)Anderson, Clark, and Mullin]{anderson1994interactive}
Anne~H Anderson, Aileen Clark, and James Mullin.
\newblock Interactive communication between children: learning how to make language work in dialogue.
\newblock \emph{Journal of Child Language}, 21\penalty0 (2):\penalty0 439--463, 1994.

\bibitem[Baker et~al.(2019)Baker, Smith, and Anissa]{baker2019educ}
Toby Baker, Laurie Smith, and Nandra Anissa.
\newblock Educ-ai-tion rebooted? exploring the future of artificial intelligence in schools and colleges, 2019.

\bibitem[Chatterjee and Bhattacharjee(2020)]{chatterjee2020adoption}
Sheshadri Chatterjee and Kalyan~Kumar Bhattacharjee.
\newblock Adoption of artificial intelligence in higher education: A quantitative analysis using structural equation modelling.
\newblock \emph{Education and Information Technologies}, 25:\penalty0 3443--3463, 2020.

\bibitem[Chen et~al.(2021)Chen, Zou, Xie, and Cheng]{chen2021twenty}
Xieling Chen, Di~Zou, Haoran Xie, and Gary Cheng.
\newblock Twenty years of personalized language learning.
\newblock \emph{Educational Technology \& Society}, 24\penalty0 (1):\penalty0 205--222, 2021.

\bibitem[Frings et~al.(2020)Frings, Hommel, Koch, Rothermund, Dignath, Giesen, Kiesel, Kunde, Mayr, Moeller, et~al.]{frings2020binding}
Christian Frings, Bernhard Hommel, Iring Koch, Klaus Rothermund, David Dignath, Carina Giesen, Andrea Kiesel, Wilfried Kunde, Susanne Mayr, Birte Moeller, et~al.
\newblock Binding and retrieval in action control (brac).
\newblock \emph{Trends in Cognitive Sciences}, 24\penalty0 (5):\penalty0 375--387, 2020.

\bibitem[Guan et~al.(2020)Guan, Mou, and Jiang]{guan2020artificial}
Chong Guan, Jian Mou, and Zhiying Jiang.
\newblock Artificial intelligence innovation in education: A twenty-year data-driven historical analysis.
\newblock \emph{International Journal of Innovation Studies}, 4\penalty0 (4):\penalty0 134--147, 2020.

\bibitem[Hwang(2003)]{hwang2003conceptual}
Gwo-Jen Hwang.
\newblock A conceptual map model for developing intelligent tutoring systems.
\newblock \emph{Computers \& Education}, 40\penalty0 (3):\penalty0 217--235, 2003.

\bibitem[Johnson et~al.(2009)Johnson, Phillips, and Chase]{johnson2009intelligent}
Benny~G Johnson, Fred Phillips, and Linda~G Chase.
\newblock An intelligent tutoring system for the accounting cycle: Enhancing textbook homework with artificial intelligence.
\newblock \emph{Journal of Accounting Education}, 27\penalty0 (1):\penalty0 30--39, 2009.

\bibitem[Lewis et~al.(2021)Lewis, Liu, Groh, and Picard]{lewis2021habit}
Robert Lewis, Yuanbo Liu, Matthew Groh, and Rosalind Picard.
\newblock Habit formation dynamics: Finding factors associated with building strong mindfulness habits.
\newblock In \emph{International Conference on Human-Computer Interaction}, pages 348--356. Springer, 2021.

\bibitem[LEWIS et~al.(2021)LEWIS, LIU, GROH, and PICARD]{lewis2021shaping}
ROBERT LEWIS, YUANBO LIU, MATTHEW GROH, and ROSALIND PICARD.
\newblock Shaping habit formation insights with shapley values: Towards an explainable ai-system for self-understanding and health behavior change.
\newblock \emph{Proc. Realizing AI Healthcare, Challenges Appearing Wild CHI}, 2021.

\bibitem[McLay et~al.(2017)McLay, Sch{\"a}fer, van~der Meer, Couper, McKenzie, O’Reilly, Lancioni, Marschik, Sigafoos, and Sutherland]{mclay2017acquisition}
Laurie McLay, Martina~CM Sch{\"a}fer, Larah van~der Meer, Llyween Couper, Emma McKenzie, Mark~F O’Reilly, Giulio~E Lancioni, Peter~B Marschik, Jeff Sigafoos, and Dean Sutherland.
\newblock Acquisition, preference and follow-up comparison across three aac modalities taught to two children with autism spectrum disorder.
\newblock \emph{International Journal of Disability, Development and Education}, 64\penalty0 (2):\penalty0 117--130, 2017.

\bibitem[Popenici and Kerr(2017)]{popenici2017exploring}
Stefan~AD Popenici and Sharon Kerr.
\newblock Exploring the impact of artificial intelligence on teaching and learning in higher education.
\newblock \emph{Research and practice in technology enhanced learning}, 12\penalty0 (1):\penalty0 22, 2017.

\bibitem[Srinivasan(2022{\natexlab{a}})]{SRINIVASAN2022100062}
Venkat Srinivasan.
\newblock Ai \& learning: A preferred future.
\newblock \emph{Computers and Education: Artificial Intelligence}, 3:\penalty0 100062, 2022{\natexlab{a}}.
\newblock ISSN 2666-920X.
\newblock \doi{https://doi.org/10.1016/j.caeai.2022.100062}.
\newblock URL \url{https://www.sciencedirect.com/science/article/pii/S2666920X22000170}.

\bibitem[Srinivasan(2022{\natexlab{b}})]{srinivasan2022ai}
Venkat Srinivasan.
\newblock Ai \& learning: A preferred future.
\newblock \emph{Computers and Education: Artificial Intelligence}, 3:\penalty0 100062, 2022{\natexlab{b}}.

\bibitem[Srinivasan and Murthy(2021)]{srinivasan2021improving}
Venkat Srinivasan and Hemavathi Murthy.
\newblock Improving reading and comprehension in k-12: Evidence from a large-scale ai technology intervention in india.
\newblock \emph{Computers and Education: Artificial Intelligence}, 2:\penalty0 100019, 2021.

\bibitem[van~der Meer et~al.(2012)van~der Meer, Sutherland, O’Reilly, Lancioni, and Sigafoos]{van2012further}
Larah van~der Meer, Dean Sutherland, Mark~F O’Reilly, Giulio~E Lancioni, and Jeff Sigafoos.
\newblock A further comparison of manual signing, picture exchange, and speech-generating devices as communication modes for children with autism spectrum disorders.
\newblock \emph{Research in Autism Spectrum Disorders}, 6\penalty0 (4):\penalty0 1247--1257, 2012.

\end{thebibliography}

\end{document}